\documentclass[sigconf, screen, nonacm]{acmart}
\usepackage[table]{xcolor} % 提供表格着色功能
\usepackage{booktabs}      % 提供 \toprule, \midrule 等高质量横线
\usepackage{multirow}
\usepackage{wasysym}
\makeatletter
\def\@affiliationfont{\small\normalfont}
\makeatother
\AtBeginDocument{%
  }
\setcopyright{none}
\begin{document}
\author{Yu Liu}
\authornote{Equal contribution.}
\affiliation{%
  \institution{Institute of Information Engineering, CAS}
  \city{Beijing}
  \country{China}}
\affiliation{%
  \institution{School of Cyber Security, UCAS}
  \city{Beijing}
  \country{China}}
\email{liuyu@iie.ac.cn}

\author{Zhiwei Yang}
\authornotemark[1]
\affiliation{%
  \institution{Institute of Information Engineering, CAS}
  \city{Beijing}
  \country{China}}
\affiliation{%
  \institution{School of Cyber Security, UCAS}
  \city{Beijing}
  \country{China}}
\email{yangzhiwei@iie.ac.cn}

\author{Wenxiao Zhang}
\authornotemark[1]
\affiliation{%
  \institution{The University of Western Australia}
  \city{Perth}
  \country{Australia}}
\email{wenxiao.zhang@research.uwa.edu.au}

\author{Cong Cao}
\authornote{Corresponding authors.}
\affiliation{%
  \institution{Institute of Information Engineering, CAS}
  \city{Beijing}
  \country{China}}
\email{caocong@iie.ac.cn}

\author{Fangfang Yuan}
\affiliation{%
  \institution{Institute of Information Engineering,CAS}
  \city{Beijing}
  \country{China}}
\email{yuanfangfang@iie.ac.cn}

\author{Kun Peng}
\affiliation{%
  \institution{Institute of Information Engineering, CAS}
  \city{Beijing}
  \country{China}}
\affiliation{%
  \institution{School of Cyber Security, UCAS}
  \city{Beijing}
  \country{China}}
\email{pengkun@iie.ac.cn}

\author{Haimei Qin}
\affiliation{%
  \institution{Institute of Information Engineering, CAS}
  \city{Beijing}
  \country{China}}
\email{qinhaimei@iie.ac.cn}

\author{Lei Jiang}
\affiliation{%
  \institution{Institute of Information Engineering, CAS}
  \city{Beijing}
  \country{China}}
\email{jianglei@iie.ac.cn}

\author{Jin B. Hong}
\affiliation{%
  \institution{The University of Western Australia}
  \city{Perth}
  \country{Australia}}
\email{jin.hong@uwa.edu.au}

\author{Hao Peng}
\affiliation{%
  \institution{Beihang University}
  \city{Beijing}
  \country{China}}
\email{penghao@buaa.edu.cn}

\author{Yanbing Liu}
\authornotemark[2]
\affiliation{%
  \institution{Institute of Information Engineering, CAS}
  \city{Beijing}
  \country{China}}
\affiliation{%
  \institution{School of Cyber Security, UCAS}
  \city{Beijing}
  \country{China}}
\email{liuyanbing@iie.ac.cn}

\renewcommand{\shortauthors}{Liu et al.}

\title{\twonotes~When the Same Musical Knowledge Forgets Differently: A Clean Probe of Pathway-Dependent Forgetting }
% \\ in Music-Capable Audio-Language Models
\begin{abstract}
A model can learn that the piano piece \emph{Für Elise} is calm and reflective by listening to the audio or by reading a text description, but does it matter which route that knowledge took when it is later at risk of being forgotten?
Forgetting research in multimodal models measures what knowledge is lost under adaptation, yet has not asked whether acquisition route affects how easily that knowledge is forgotten.
We call this untested premise the \textit{Pathway-Invariant Assumption}.
Music understanding enables a clean test because a music clip and a canonical text description can be aligned to the same perceptual content, allowing the same knowledge unit to enter a model through listening or reading while the target remains fixed.
Across multiple architecturally distinct audio-language models, we observe a consistent asymmetry: text-pathway knowledge is forgotten more than matched audio-pathway knowledge under identical adaptation pressure.
To attribute this effect to route rather than confounds, we introduce the \textbf{P}aired \textbf{P}athway \textbf{C}ontrolled \textbf{P}rotocol (\textbf{PPCP}), a three-phase design that establishes matched pathway baselines, activates both pathways under symmetric supervision on the same knowledge pool, and applies identical forgetting pressure to both pathways.
The gap is stable across models and gain-controlled analyses, persists when contradictory overwrite is replaced by correct-label cross-domain learning, remains under single-modality pressure, and is not removed by lightweight replay.
Two independent routing-depth controls confirm that the effect is not explained by architectural depth, pointing to input representation as the dominant factor.
Under PPCP, our results demonstrate that forgetting is highly route-dependent, establishing acquisition route as a new analytical dimension for forgetting research and multimodal system design.
\par\vskip\smallskipamount\relax
\noindent\textbf{Code:} \url{https://github.com/Ameame1/Audio_Memory_PPCP}
% Under PPCP, these results suggest that retention depends not only on what a model learned, but also on how that knowledge entered the model, establishing acquisition route as a new analytical dimension for forgetting research and multimodal system design.
\end{abstract}
% \ccsdesc[500]{Information systems~Music retrieval}
% \ccsdesc[300]{Computing methodologies~Natural language processing}
\maketitle

\begin{figure}[t]
\centering
\includegraphics[width=\linewidth]{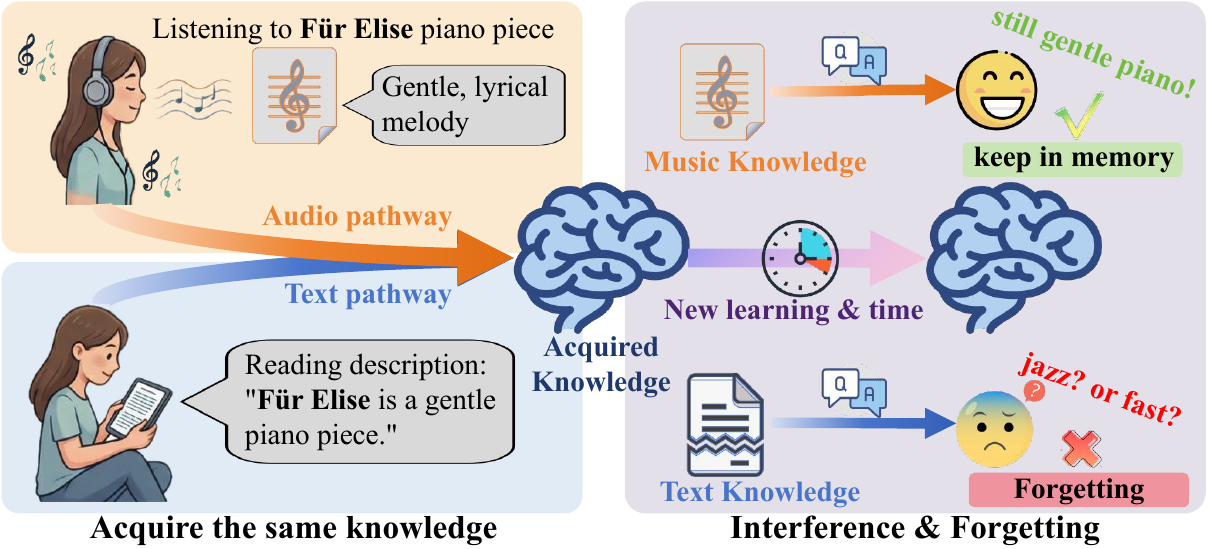}
\Description{Schematic comparing audio and text acquisition pathways for the same musical knowledge under later interference.}
% \caption{A schematic illustration of path-dependent forgetting based on human memory mechanisms. For the same musical knowledge, memories formed through an audio-listening path are more resistant to interference than those formed through a text-reading path.}
\caption{Schematic of path-dependent forgetting: for the same musical knowledge, audio-based memories are more resistant to interference than text-based memories.}
\label{toy}
\par\vskip-2mm\relax
\end{figure}

%%
%% ─── INTRODUCTION ───────────────────────────────────────────────────────────
%%

\section{Introduction}

Consider a question asked of an audio-language model: \emph{what is the mood of Für Elise?}
Unlike a speech recognition task, where the answer can only be derived from the acoustic signal itself, this question admits two distinct routes.
The model may arrive at the answer by processing the raw waveform, attending to the gentle dynamics and lyrical phrasing of the piano.
Alternatively, it may arrive at the same answer by reading a textual description of the same piece, one that captures its calm, reflective character in language.
Both routes lead to the same musical knowledge, yet they engage fundamentally different computational pathways within the same model.
Music understanding makes this dual-pathway structure especially clean: the answer-relevant attributes of a music recording (genre, instrumentation, mood, texture) are categorical and well-defined, and a canonical textual description can be constructed to capture them with sufficient fidelity for the question-answer categories studied.
This makes music the first domain where a fundamental question can be posed under proper experimental control: once the same knowledge has entered a model through different pathways, is it equally vulnerable to later forgetting?

This question has never been asked, for two reasons.
First, existing forgetting research, from continual learning~\cite{vandeven2024continual,liu2025continual,wei2025mitigating} to model editing~\cite{meng2022rome,chen2025lifelong,zhangreliable,thede2025wikibigedit} to multimodal adaptation~\cite{zhai2024investigating}, operates at the level of tasks, datasets, or modalities as a whole; acquisition route is never tracked as a variable.
We call this the \emph{Pathway-Invariant Assumption}.
Second, even if one tried, continuation data whose modality composition is uncontrolled produces gradient updates that are never pathway-neutral, making any observed asymmetry impossible to attribute to pathway identity alone.
We call this the \emph{Symmetric Continuation Bias}.
Together, these two blind spots have rendered pathway-dependent retention invisible to existing frameworks.

Across multiple architecturally distinct audio-language models, we find controlled evidence against the Pathway-Invariant Assumption: text-pathway knowledge is consistently forgotten more than matched audio-pathway knowledge, a phenomenon we term \emph{pathway-dependent forgetting}.
To attribute this effect to acquisition route rather than confounds, we introduce the Paired Pathway Controlled Protocol (PPCP), a three-phase framework that isolates pathway as the primary variable under investigation.
We then ask \emph{why} the gap exists: it persists when text-side pressure is removed, when the audio projector is perturbed, when contradictory replacement is substituted with correct-label cross-domain learning, when routing depth is controlled via two independent probes, when the implicated decoder layers are frozen, and when lightweight replay is applied.
Six controls, one consistent answer: the asymmetry is structural, tied to the nature of input representation rather than any single architectural or training factor (Table~\ref{tab:roadmap}).

Across all evaluated models, the same musical fact is consistently more resistant to forgetting when acquired by listening than when acquired by reading, and this pathway gap persists across all six controls.
Together, these results indicate that acquisition pathway is a first-class variable in forgetting, alongside content and adaptation pressure.
Practically, this implies that forgetting interventions should be pathway-aware: continual learning may require pathway-specific resource allocation, and model editing or unlearning should not assume that changes made through one pathway will automatically transfer to the other.
Our contributions are as follows:

\begin{enumerate}
    \item \textbf{A new problem formulation.}
    We identify two blind spots in existing forgetting research, the \emph{Pathway-Invariant Assumption} and the \emph{Symmetric Continuation Bias}, and formalize \emph{pathway-dependent forgetting} as a testable phenomenon: does the route through which knowledge enters a multimodal model shape its vulnerability to later forgetting?

    \item \textbf{A clean paired protocol.}
    We propose PPCP, the first experimental framework that jointly enforces target equivalence, symmetric supervision, leakage control, and acquisition comparability across aligned pathways.

    \item Under PPCP, across four architecturally distinct audio language models, text-pathway knowledge is consistently forgotten more than matched audio-pathway knowledge, and six targeted controls rule out all tested alternative explanations.
\end{enumerate}

\section{Related Work}
\label{sec:related}

\subsection{Catastrophic Forgetting}

Catastrophic forgetting has been a central concern since the early connectionist literature~\cite{mccloskey1989catastrophic,french1999catastrophic}, and continual learning has developed a broad family of mitigation strategies including replay~\cite{chaudhry2019continual}, parameter regularization~\cite{kirkpatrick2017overcoming}, and architectural isolation~\cite{rusu2016progressive}; see~\cite{vandeven2024continual,wang2024comprehensive} for surveys. With the rise of large language models, forgetting has re-emerged in continual fine-tuning and instruction tuning~\cite{wei2022finetuned}, where adaptation on new data can degrade previously acquired capabilities~\cite{shi2024continualllm}. A related line of work in model editing and machine unlearning~\cite{maini2024tofu,meng2022rome,mitchell2022memit} focuses on removing or rewriting target facts while preserving locality. Across all of these frameworks, the unit of analysis is the task, the dataset, or the individual fact. What none of them track is the \emph{route} through which a knowledge unit was originally acquired: retention is implicitly treated as depending on content and adaptation pressure alone, with acquisition pathway absent as an experimental variable.

\subsection{Multimodal Forgetting and Understanding}
Recent work has shown that integrating non-text encoders into language models can degrade existing capabilities~\cite{zhai2024investigating,zheng2024mllm_cl}. However, these studies treat the modality or task as the unit of analysis, asking whether adding a new modality harms performance on an existing one. Modality-inconsistent continual learning~\cite{pian2024micl,tang2024mitigating} is related, but it changes both task content and input modality across training stages, preventing attribution of forgetting to the acquisition route alone. In contrast, we fix both the knowledge unit and the model, and vary the acquisition pathway as the primary factor.
In music, audio-language models are beginning to scale toward richer understanding~\cite{ghosh2025musicflamingo,pan2025almssurvey}, but prior work remains centered on speech tasks such as recognition, translation, and dialogue. Retention of musical knowledge under adaptation has not been studied. More broadly, no existing protocol controls the modality composition of continuation data when measuring retention within a single model. We formalize this gap as \emph{Symmetric Continuation Bias} in Section~\ref{sec:prelim}.

%%
%% ─── BLIND SPOTS ────────────────────────────────────────────────────────────

\begin{figure*}[t!]
\centering
\includegraphics[width=\linewidth]{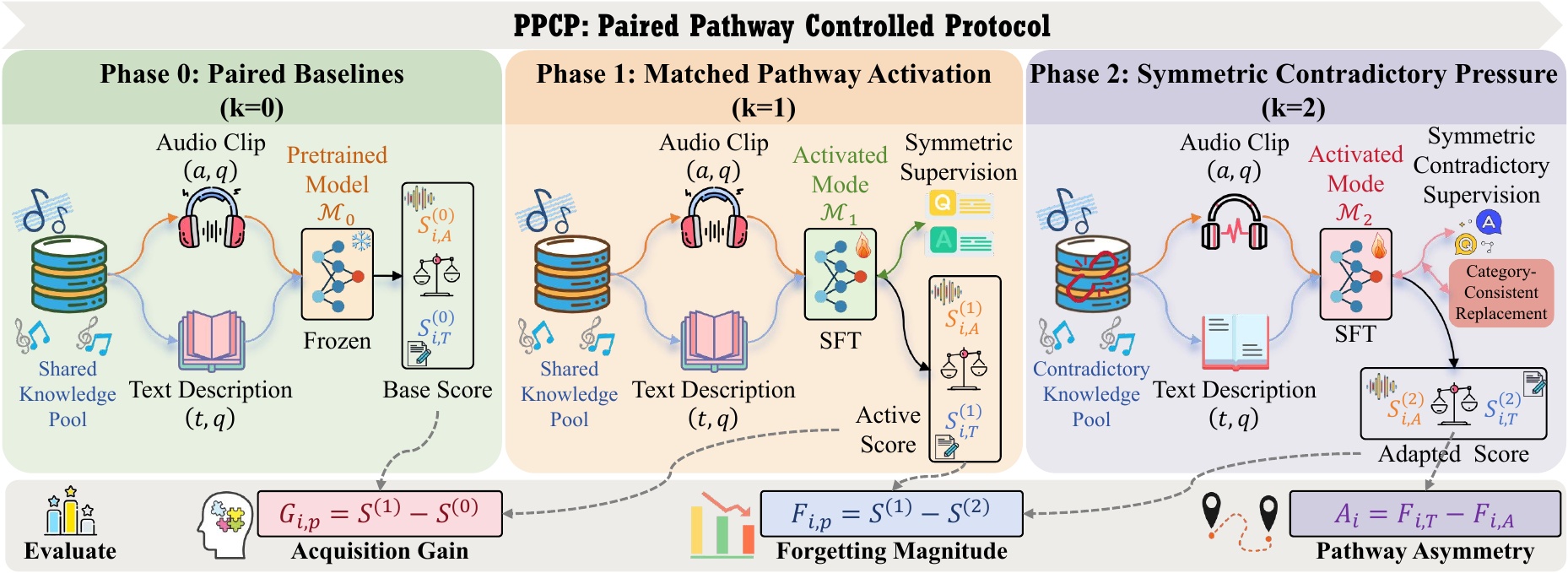}
\Description{Three-phase PPCP workflow showing paired baselines, matched pathway activation, and symmetric contradictory pressure for audio and text pathways.}
\caption{Overview of PPCP. The same musical knowledge unit is presented to both the audio pathway (A2T) and the text pathway (T2T) under matched supervision in Phase~1, and both pathways receive identical contradictory supervision in Phase~2. Per-item scores are recorded at every checkpoint, making acquisition pathway the primary experimental variable.}
\label{fig:overview}
\end{figure*}
\section{Blind Spots in Current Forgetting}
\label{sec:prelim}
We identify two structural blind spots that prevent existing forgetting frameworks from directly testing whether acquisition route affects retention in multimodal models.

\textit{The Pathway-Invariant Assumption.}
Existing frameworks implicitly assume that retention is independent of how knowledge enters the model.
Whether they estimate parameter importance~\cite{kirkpatrick2017overcoming}, edit factual associations~\cite{maini2024tofu,meng2022rome}, or measure performance across sequential stages~\cite{vandeven2024continual,wang2024comprehensive}, the unit of analysis is always the task, fact, or dataset---not the acquisition route of a specific knowledge item.
This is largely innocuous in unimodal settings, where pathway identity is fixed.
In multimodal models, however, the same knowledge can enter through multiple aligned pathways, and music makes this particularly clear: the attributes we study (genre, instrumentation, mood, texture) are categorical and can be described textually with sufficient fidelity for controlled pathway-level comparison.

\textit{The Symmetric Continuation Bias.}
Under continued training, the modality composition of continuation data determines which pathways receive the strongest updates: audio-heavy data mainly updates the audio encoder and projector, while text-heavy data primarily affects language-side representations.
Neither provides a pathway-neutral probe of forgetting.
Prior multimodal forgetting studies~\cite{zhai2024investigating,zheng2024mllm_cl,huai2025cl} and modality-inconsistent continual learning~\cite{pian2024micl,tang2024mitigating} vary task and modality together, so they compare modalities or tasks as wholes rather than matched acquisition routes for the same knowledge unit.
To our knowledge, no existing protocol equalizes forgetting pressure across pathways.
Together, these two blind spots motivate the paired, pathway-controlled protocol introduced next.

%%
%% ─── METHODOLOGY ────────────────────────────────────────────────────────────
%%

\section{Methodology}
\label{sec:method}

\subsection{Problem Statement}
\label{sec:problem}

We investigate whether the same musical knowledge, once acquired through different aligned pathways within a single audio-language model, is equally vulnerable to subsequent forgetting, i.e., a within-model comparison of two aligned access routes to one knowledge unit. Because a direct pathway comparison is confounded without controls on differential acquisition strength and overwrite symmetry, we explicitly account for both. Let $\mathcal{M}_{\theta}$ denote an audio-language model with parameters $\theta$, and for each knowledge unit~$i$ define $x_i=(a_i,\, t_i,\, q_i,\, y_i,\, c_i)$, where $a_i$ is an audio clip, $t_i$ a semantically aligned textual description, $q_i$ a natural-language question, $y_i$ the target answer, and $c_i$ the set of category-consistent candidates containing $y_i$ (\emph{e.g.}, all genre labels when $q_i$ asks about genre). The same unit is probed through two pathways: an \textbf{audio pathway} (A2T) with input $(a_i, q_i)$ and a \textbf{text pathway} (T2T) with input $(t_i, q_i)$, both evaluated against the same target $y_i$; under identical adaptation pressure, we ask whether retention through A2T and T2T degrades symmetrically or asymmetrically.

\begin{center}
\fbox{%
\begin{minipage}{0.97\linewidth}
\textbf{Protocol 1: Paired Pathway Controlled Protocol (PPCP)}\\[0.2em]
\textbf{Input:} pretrained model $\mathcal{M}_{\theta_0}$, paired knowledge pool $\mathcal{K}$, held-out paired evaluation set $\mathcal{E}$.\\
\textbf{Preprocessing (Leakage control).} Build canonical text descriptions and apply leakage filtering before Phase~1, as detailed in Section~\ref{sec:setup}. \hfill\textit{(satisfies P3)}\\[0.3em]
\textbf{Phase~0 (Paired baselines).}
Evaluate $\mathcal{M}_{\theta_0}$ on $\mathcal{E}$ through both pathways, yielding per-item baseline scores $S^{(0)}_{i,A}$ and $S^{(0)}_{i,T}$.\\[0.3em]
\textbf{Phase~1 (Matched pathway activation).}
Starting from $\mathcal{M}_{\theta_0}$, fine-tune on a balanced mixture of A2T and T2T examples drawn from $\mathcal{K} \setminus \mathcal{E}$, together with a small amount of general instruction data, yielding $\mathcal{M}_{\theta_1}$; record $S^{(1)}_{i,A}$ and $S^{(1)}_{i,T}$ on $\mathcal{E}$. \hfill\textit{(satisfies P1, P4)}\\[0.3em]
\textbf{Phase~2 (Symmetric contradictory pressure).}
Starting from $\mathcal{M}_{\theta_1}$, for each training item $i \in \mathcal{K} \setminus \mathcal{E}$, replace the original answer $y_i$ with a contradictory substitute $\tilde{y}_i$ drawn uniformly from $c_i \setminus \{y_i\}$ (i.e., a different answer from the same category set).
Fine-tune on both-pathway contradictory data using these substituted labels, yielding $\mathcal{M}_{\theta_2}$; record $S^{(2)}_{i,A}$ and $S^{(2)}_{i,T}$ on $\mathcal{E}$. \hfill\textit{(satisfies P2)}\\[0.3em]
\textbf{Output:} per-item score triples $\{S^{(k)}_{i,p}\}_{k=0,1,2}$ for both pathways.
\end{minipage}%%
}
\end{center}

\subsection{Structural Prerequisites for Pathway-Level Comparison}
\label{sec:requirements}

We identify four conditions that any valid pathway-level forgetting comparison must satisfy.
To our knowledge, no existing protocol jointly satisfies all of them.

\noindent$\bullet$ \textbf{(P1) Target equivalence.}
Both pathways must correspond to the same knowledge unit, so that any retention difference reflects \emph{how} knowledge was acquired rather than \emph{what} was acquired.

\noindent$\bullet$ \textbf{(P2) Shared and symmetric supervision.}
Both pathways must receive identical overwrite pressure during the forgetting phase; modality-biased continuation data would confound the comparison.

\noindent$\bullet$ \textbf{(P3) No trivial leakage.}
The text description $t_i$ must not expose the answer $y_i$ in a way that makes T2T trivially easier for reasons unrelated to pathway structure.

\noindent$\bullet$ \textbf{(P4) Acquisition comparability.}
If one pathway systematically learns more strongly during the acquisition phase, higher forgetting on that pathway may be an artifact of higher acquisition rather than weaker retention.
The protocol must record per-item acquisition gain and provide gain-controlled analysis.
\par\vskip-0.2cm\relax
\subsection{The Paired Pathway Controlled Protocol (PPCP)}
\label{sec:protocol}

PPCP is a three-phase framework that satisfies all four prerequisites above.
Figure~\ref{fig:overview} illustrates its structure.
Let $\mathcal{K}$ denote the full paired knowledge pool.
Let $\mathcal{E} \subset \mathcal{K}$ denote the held-out paired evaluation set, disjoint from the training split.
Phases~1 and~2 train on $\mathcal{K} \setminus \mathcal{E}$, while pathway-level scores are recorded on $\mathcal{E}$ at every checkpoint.

\noindent\textbf{Notation.} $i$ indexes paired knowledge units in $\mathcal{K}$; $p \in \{A,T\}$ denotes the audio or text pathway; $k \in \{0,1,2\}$ denotes the checkpoint after Phase~$k$; $S^{(k)}_{i,p}$ is a scalar score measuring how well the model's response on item $i$ through pathway $p$ matches the reference answer $y_i$ at checkpoint $k$ (concrete instantiations are given in Section~\ref{sec:evalframework}).

\textbf{Phase~0} provides paired pathway baselines before any controlled adaptation.

\textbf{Phase~1} does not primarily inject new knowledge, because pretrained models already contain substantial musical understanding.
Instead, it establishes a controlled checkpoint where both pathways are actively engaged under matched supervision on the same knowledge pool, satisfying \emph{target equivalence}~(P1) and \emph{acquisition comparability}~(P4).

\textbf{Phase~2} keeps substitutes semantically plausible via category-consistent replacement while maximizing overwrite pressure.
Both pathways receive the same $\tilde{y}_i$ for each item, enforcing \emph{shared and symmetric supervision}~(P2).
A potential concern is that audio-conditioned representations may resist substitution more strongly than text-side representations; we address this through controlled experiments in Section~\ref{sec:rq2} and formalize the resulting pathway--depth entanglement in Section~\ref{sec:entanglement}.
Prerequisite~(P3) (no trivial leakage) is enforced during data construction and filtering, detailed in Section~\ref{sec:setup}.

\noindent\textbf{Training.} Both phases use supervised fine-tuning (SFT) with mixed precision and gradient accumulation. Checkpoints are saved periodically during Phase~2 for trajectory analysis. Model-specific training configurations are detailed in Section~\ref{sec:setup}.
\begin{table*}[t]
\centering
\small
\caption{Experimental roadmap: each candidate explanation for the observed asymmetry, the control designed to test it, and the outcome. The rightmost column indicates whether the explanation is ruled out as sufficient.}
\label{tab:roadmap}
\begin{tabular}{lllc}
\toprule
\textbf{Candidate Explanation} & \textbf{Testable Prediction} & \textbf{Control Experiment} & \textbf{Ruled Out?} \\
\midrule
Differential acquisition      & T2T has more to lose from Phase\,1          & (\textbf{Main \& PPCP}) ANCOVA on $G_{i,p}$   (Table~\ref{tab:crossmodel}, RQ1)              & \checkmark \\
Direct text overwrite          & Removing text pressure closes gap            & \textbf{(a)}\,Audio-only forgetting (Table~\ref{tab:rq2}\,a; RQ2)                 & \checkmark \\
Projector shielding            & Perturbing projector reduces gap             & \textbf{(b)}\,Gaussian noise $\sigma{\in}\{0.50\text{--}1.00\}$ (Table~\ref{tab:rq2}\,b; RQ2)  & \checkmark \\
Contradictory replacement      & Correct labels eliminate gap                 & \textbf{(c)}\, MELD emotion classif.\ (Table~\ref{tab:rq2}\,c; RQ2)           & \checkmark \\
Routing depth protection       & Extra routing replicates gap  & \textbf{(d)}\,TTS + text-routed probes (Table~\ref{tab:rq2}\,d; RQ2)       & \checkmark \\
Layer-specific gradient pressure & Freezing T2T-dominant layers compresses gap  & \textbf{(e)}\,Freeze layers 0--8\,/\,10--29 (Table~\ref{tab:rq2}\,e; RQ2)         & \checkmark \\
Shallow training artifact      & Replay closes the gap                       & \textbf{(f)}\,Balanced \& T2T-targeted replay (Table~\ref{tab:replay}; RQ3)             & \checkmark \\
\bottomrule
\end{tabular}

\end{table*}
\begin{table*}[!t]
\centering
\small
\caption{Pathway asymmetry across all evaluated runs. All models show that the text pathway (T2T) forgets significantly more than the audio pathway (A2T). All bootstrap 95\% CI lower bounds remain strictly above zero, and all $p$-values are $< 0.001$.}
\label{tab:crossmodel}
\resizebox{\textwidth}{!}{
\begin{tabular}{l cc c cc ccccc}
\toprule
& \multicolumn{2}{c}{\textbf{Raw Forgetting}} & \multicolumn{3}{c}{\textbf{Pathway Asymmetry}} & \multicolumn{5}{c}{\textbf{Statistical Analysis}} \\
\cmidrule(lr){2-3} \cmidrule(lr){4-6} \cmidrule(lr){7-11}
Model & $D_{\text{A2T}}$$\downarrow$ & $D_{\text{T2T}}$$\downarrow$ & $\Delta D^{\text{gen}}$$\uparrow$ & RRS$^{\text{gen}}$$\uparrow$ & RRS$^{\text{logit}}$$\uparrow$ & $d$$\uparrow$ & $p$$\downarrow$ & $\Delta D^{\text{gen}}$ CI & $\Delta D^{\text{logit}}$ CI & RRS$^{\text{gen}}$ CI \\
\midrule
Qwen2-Audio$_{\text{seed42}}$~\cite{chu2024qwen2audio} & 0.235 & \textbf{0.289} & \textbf{+0.054} & +0.187 & +0.402 & 0.289 & $<0.001$ & $[+.038,\,+.071]$ & $[+.267,\,+.367]$ & $[+.133,\,+.244]$ \\
Qwen2-Audio$_{\text{seed43}}$~\cite{chu2024qwen2audio} & 0.233 & \textbf{0.287} & \textbf{+0.054} & +0.189 & +0.403 & 0.297 & $<0.001$ & $[+.038,\,+.070]$ & $[+.263,\,+.360]$ & $[+.135,\,+.244]$ \\
Qwen2-Audio$_{\text{seed44}}$~\cite{chu2024qwen2audio} & 0.228 & \textbf{0.281} & \textbf{+0.053} & +0.188 & +0.423 & 0.282 & $<0.001$ & $[+.036,\,+.069]$ & $[+.278,\,+.377]$ & $[+.130,\,+.245]$ \\
% \addlinespace % 在视觉上稍微隔开
SALMONN~\cite{tang2024salmonn} & 0.232 & \textbf{0.276} & \textbf{+0.044} & +0.158 & +0.289 & 0.288 & $<0.001$ & $[+.031,\,+.057]$ & $[+.177,\,+.232]$ & $[+.112,\,+.206]$ \\
Audio Flamingo~3~\cite{goel2025af3} & 0.175 & \textbf{0.281} & \textbf{+0.106} & +0.377 & +0.302 & 0.517 & $<0.001$ & $[+.088,\,+.124]$ & $[+.199,\,+.262]$ & $[+.309,\,+.449]$ \\
Qwen2.5-Omni~\cite{xu2025qwen25omni} & 0.242 & \textbf{0.271} & \textbf{+0.029} & +0.107 & +0.358 & 0.213 & $<0.001$ & $[+.017,\,+.041]$ & $[+.185,\,+.273]$ & $[+.064,\,+.150]$ \\
\bottomrule
\end{tabular}%
}
\end{table*}
\subsection{Evaluation Framework}
\label{sec:evalframework}

Let $S^{(k)}_{i,p}$ denote the score measured through pathway $p \in \{A,T\}$ at checkpoint $k \in \{0,1,2\}$ for item $i \in \mathcal{E}$.
We instantiate $S^{(k)}_{i,p}$ in two complementary views:

\begin{itemize}
    \item \emph{Generation space}: $S^{(k)}_{i,p}$ is the cosine similarity between sentence embeddings~\cite{reimers2019sentencebert} of the model's generated answer and the reference answer $y_i$.
    \item \emph{Logit space}: $S^{(k)}_{i,p}$ is the mean log-probability of reference answer tokens under teacher forcing.
\end{itemize}

\noindent The two views are complementary: generation-space forgetting can understate the underlying retention shift, while logit-space forgetting can reveal asymmetries that surface-level outputs attenuate.

\noindent$\bullet$ \textbf{Item-level quantities.}
For each item $i$, the \emph{acquisition gain} $G_{i,p}$ captures how much each pathway learned during Phase~1:
\begin{equation}
G_{i,p} = S^{(1)}_{i,p} - S^{(0)}_{i,p}.
\label{eq:gain}
\end{equation}
The \emph{forgetting magnitude} $F_{i,p}$ captures how much of that knowledge is lost after Phase~2:
\begin{equation}
F_{i,p} = S^{(1)}_{i,p} - S^{(2)}_{i,p}.
\label{eq:forget}
\end{equation}
A positive $F_{i,p}$ indicates that pathway $p$ has lost knowledge on item $i$ during Phase~2.
The \emph{per-item asymmetry} $A_i$ compares forgetting across pathways on the same item:
\begin{equation}
A_i = F_{i,T} - F_{i,A}.
\label{eq:asym}
\end{equation}
A positive $A_i$ means that text-pathway knowledge on item $i$ is more severely degraded than the matched audio-pathway knowledge.

\noindent$\bullet$ \textbf{Aggregate metrics.}
Let $D_p$ denote the mean forgetting on pathway $p$:
\begin{equation}
D_p = \frac{1}{|\mathcal{E}|}\sum_{i \in \mathcal{E}} F_{i,p}.
\label{eq:Dp}
\end{equation}
The \emph{absolute forgetting gap} $\Delta D$ measures how much more the text pathway forgets than the audio pathway on average:
\begin{equation}
\Delta D = D_T - D_A = \frac{1}{|\mathcal{E}|}\sum_{i \in \mathcal{E}} A_i.
\label{eq:deltaD}
\end{equation}
To express this gap as a proportion of text-pathway forgetting, we define the \emph{relative retention shift} (RRS):
\begin{equation}
\mathrm{RRS} = \frac{D_T - D_A}{D_T}.
\label{eq:rrs}
\end{equation}
Intuitively, RRS expresses how much of the text pathway's total forgetting is \emph{excess} degradation not shared by the audio pathway; a higher value indicates a larger structural disadvantage for the text pathway.

\noindent$\bullet$ \textbf{Hypothesis testing.}
The null hypothesis posits that the expected item-level asymmetry is zero; the alternative posits a systematic positive shift:
\begin{equation}
H_0\!: \mathbb{E}[A_i] = 0, \qquad H_1\!: \mathbb{E}[A_i] > 0.
\end{equation}
Rejection of $H_0$ with positive $\Delta D$ constitutes evidence for \emph{pathway-dependent forgetting}.
We evaluate this via one-tailed Wilcoxon signed-rank tests~\cite{wilcoxon1945individual} on the paired $\{A_i\}$ distribution and report Cohen's $d = \bar{A}/s_A$~\cite{cohen1988statistical} as the standardized effect size.
Bootstrap confidence intervals~\cite{efron1979bootstrap} are computed for both $\Delta D$ and RRS; the confidence level and number of resamples are specified in Section~\ref{sec:setup}.

\noindent$\bullet$ \textbf{Acquisition confound control.}
Higher forgetting on the text pathway could in principle reflect stronger Phase~1 acquisition rather than weaker retention: a pathway that learned more has more to lose, so observing larger $F_{i,T}$ does not by itself imply greater structural vulnerability.
To disentangle acquisition strength from pathway identity, we use analysis of covariance (ANCOVA)~\cite{cochran1957ancova}, a linear model that combines the logic of ANOVA---testing whether a categorical factor (here, pathway) affects the outcome---with regression---adjusting for a continuous covariate (here, Phase~1 acquisition gain).
Concretely, we model forgetting as
\begin{equation}
F_{i,p} = \mu + \beta_{\text{path}} \cdot \mathbf{1}[p = T] + \beta_{\text{gain}} \cdot G_{i,p} + \epsilon_{i,p},
\label{eq:ancova}
\end{equation}
where $\mu$ is the intercept,
$\mathbf{1}[p = T]$ is the indicator variable that equals 1 when the pathway is text and 0 otherwise,
$\beta_{\text{gain}}$ absorbs the linear relationship between how much was learned ($G_{i,p}$) and how much was forgotten ($F_{i,p}$),
$\beta_{\text{path}}$ captures the residual pathway effect on forgetting \emph{after} this acquisition-driven component has been removed,
and $\epsilon_{i,p}$ is the residual error term.
A significantly positive $\beta_{\text{path}} > 0$ would indicate that the text pathway forgets more even among items with comparable acquisition strength, constituting evidence that the asymmetry is a property of the pathway itself rather than an artifact of differential learning.
% ─── EXPERIMENTS ────────────────────────────────────────────────────────────
\section{Experiments}
\label{sec:experiments}
We structure our experimental investigation around three research questions; all experiments are conducted under \textbf{PPCP}, with a full roadmap of candidate explanations and controlled experiments given in Table~\ref{tab:roadmap}.
\textbf{RQ1:} Does pathway-dependent forgetting exist, and is it robust across architectures, seeds, metrics, and question categories?
\textbf{RQ2:} What Drives the Asymmetry?
\textbf{RQ3:} Can the Asymmetry Be Mitigated?

\subsection{Experimental Setup}
\label{sec:setup}

\noindent$\bullet$ \textbf{Data.}
We use MusicQA~\cite{liu2024mullama}, which provides open-ended QA pairs over music audio across genre, instrument, mood, tempo, vocal, and sound-description categories. Following Section~\ref{sec:protocol}, we derive one canonical description per clip from the ``Describe the audio'' answer, restrict to non-numeric, non-temporal content, remove the source question and two high-overlap families identified by embedding-similarity audit, and enforce audio-level train/test separation. After filtering, Phase~1 comprises 6{,}750 A2T + 6{,}750 T2T examples from a shared knowledge pool plus 1{,}500 general instructions; Phase~2 comprises 12{,}000 A2T + 12{,}000 T2T examples with category-consistent replacement answers; the test set contains 500 held-out items. Additional controls are detailed in Section~\ref{sec:meld} (cross-domain MELD~\cite{poria2019meld} control) and Section~\ref{sec:disentangle} (routing-depth probes).

\noindent$\bullet$ \textbf{Models.}
We evaluate four music-capable audio-language models spanning distinct architectural paradigms: Qwen2-Audio-7B-Instruct~\cite{chu2024qwen2audio}, SALMONN~\cite{tang2024salmonn}, Audio Flamingo~3 (AF3)~\cite{goel2025af3}, and Qwen2.5-Omni~\cite{xu2025qwen25omni}.
These models differ in encoder design (Whisper~\cite{radford2023whisper}, BEATs~\cite{chen2023beats}+Whisper, AF-Whisper), projection mechanism (linear, Q-Former~\cite{li2023blip2}, MLP, none), language model backbone (Qwen
2-7B, Vicuna~\cite{chiang2023vicuna}, Qwen-2.5-7B), and overall paradigm (encoder--projector--LLM vs.\ Thinker--Talker).

\noindent$\bullet$ \textbf{Evaluation metrics.}
\label{sec:metrics}
Aggregate forgetting quantities are defined in Section~\ref{sec:evalframework}.
The primary generation-space metric uses all-MiniLM-L6-v2~\cite{reimers2019sentencebert}; logit-space RRS serves as a secondary diagnostic.
Across \textbf{RQ1}--\textbf{RQ3}, our statistical toolkit includes paired significance testing, standardized effect sizes, interval estimation, acquisition-confound adjustment, and leakage sanity checks.
Natural Language Inference (NLI) entailment scores and BERTScore~\cite{zhang2020bertscore} are computed as additional robustness checks.
Statistical significance is assessed via one-tailed Wilcoxon signed-rank tests with Cohen's $d$ as effect size; bootstrap confidence intervals (95\%, 10{,}000 resamples) are reported for $\Delta D$ and RRS; acquisition confounds are controlled via ANCOVA (Eq.~\ref{eq:ancova}); and per-item leakage is checked with Spearman rank correlation.

\noindent$\bullet$ \textbf{Training details.}
All models are trained with bfloat16 precision using gradient accumulation to maintain a consistent effective batch size.
Phase~1 runs for 3 epochs and Phase~2 for 2 epochs across all models.
For Qwen2-Audio~\cite{chu2024qwen2audio}, we perform full fine-tuning and run three independent seeds (42, 43, 44) for cross-seed stability analysis.
Audio Flamingo~3~\cite{goel2025af3} uses a custom HuggingFace trainer due to framework incompatibility.
SALMONN~\cite{tang2024salmonn} freezes its dual audio encoder and Q-Former~\cite{li2023blip2} bridge, fine-tuning only the Vicuna~\cite{chiang2023vicuna} backbone.
Qwen2.5-Omni~\cite{xu2025qwen25omni} applies full fine-tuning to its Thinker module.

\noindent$\bullet$ \textbf{Controlled experiments overview.}
Beyond the primary four-model evaluation, we conduct six targeted controls on Qwen2-Audio (seed~42) to rule out alternative explanations:
\textbf{(a)}~\emph{Pressure-modality variation}: Phase~2 uses only A2T items (audio-only) or only T2T items (text-only), testing whether direct overwriting of text representations is necessary.
\textbf{(b)}~\emph{Projector perturbation}: Gaussian noise ($\sigma \in \{0.50, 0.75, 1.00\}$) is injected into audio projector weights after Phase~1, with a decoder-noise control, testing whether projector geometry protects the audio pathway.
\textbf{(c)}~\emph{MELD cross-domain control}: Phase~2 is replaced by correct-label emotion classification from MELD~\cite{poria2019meld} (12{,}000 items per arm, paired audio and transcript with shared emotion labels), testing whether contradictory replacement is necessary.
\textbf{(d)}~\emph{Routing-depth probes} (two probes used jointly): a TTS-routed probe feeds synthesized speech through the full audio stack, and a text-routed probe adds an extra linear block to the text pathway; the two have orthogonal confound structures and jointly test whether architectural depth explains the asymmetry.
\textbf{(e)}~\emph{Layer freezing}: decoder layers 0--8 or 10--29 are frozen during Phase~2, testing whether the gradient distribution pattern is causally responsible.
\textbf{(f)}~\emph{Replay}: balanced (5\%, 10\%) and T2T-targeted (5\%) replay are mixed into Phase~2, testing whether lightweight mitigation closes the gap.
Table~\ref{tab:roadmap} maps each candidate explanation to its control; Table~\ref{tab:rq2} reports all results; \textit{full hyperparameters}, \textit{per-category statistics}, \textit{metric robustness data}, \textit{additional control details}, and \textit{baseline scores} are in \textbf{Appendices A--E}.
\par\vskip-0.4cm\relax
\begin{figure*}[!t]
\centering
\includegraphics[width=0.93\linewidth]{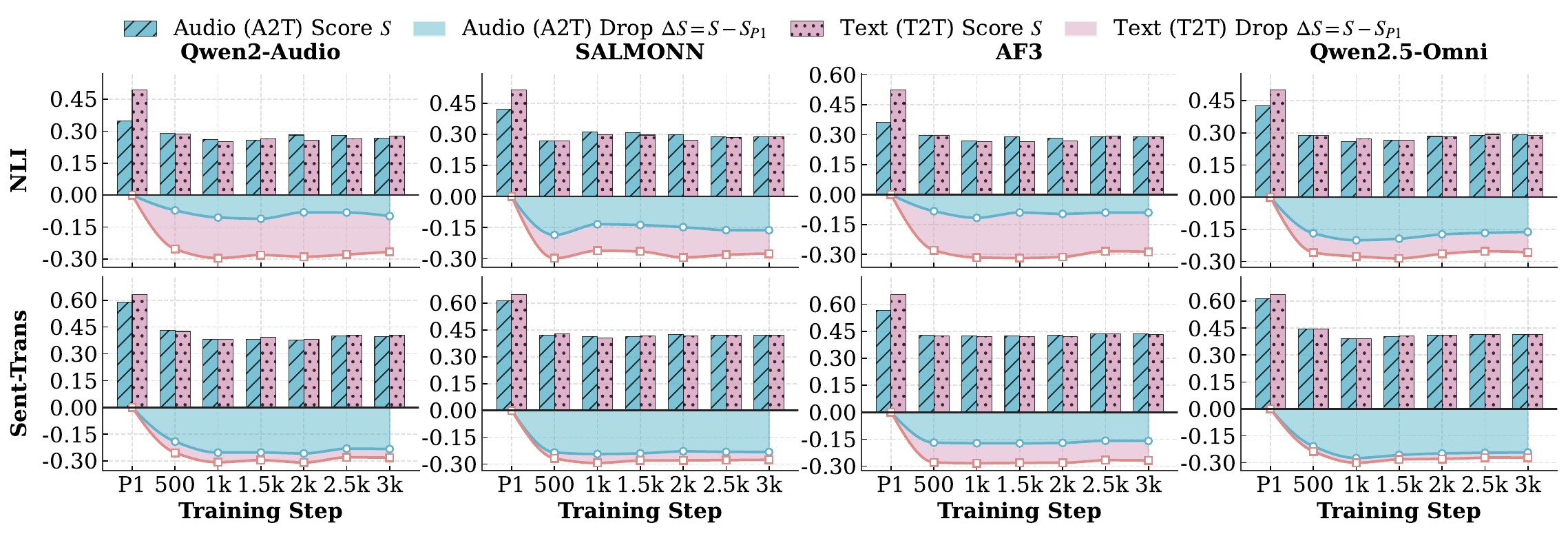}
\Description{Metric robustness figure comparing pathway performance and degradation across models and training steps.}
\par\vskip-3mm\relax
\caption{Trajectory of modality-specific performance and degradation across training steps, where the upper bars represent absolute scores ($S$) and the lower shaded areas indicate relative performance drops ($\Delta S = S - S_{P1}$) from the baseline (Phase 1).}
\label{fig:metric_heatmap}
\end{figure*}

\begin{figure}[t]
\centering
\includegraphics[width=\linewidth]{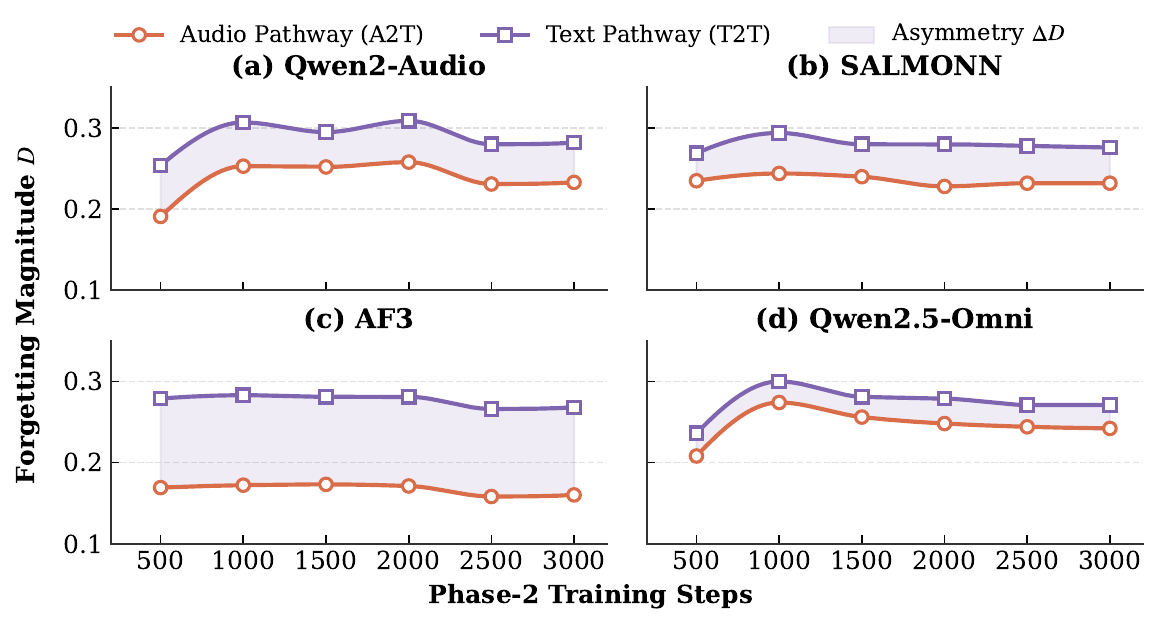}
\Description{Line plots of A2T and T2T forgetting magnitude during Phase 2 across four models.}
\par\vskip-3mm\relax
\caption{Per-pathway forgetting trajectory during Phase~2. The text pathway (orange) forgets more than the audio pathway (blue) from the earliest checkpoint onward.}
\label{fig:trajectory}
\par\vskip-0.5cm\relax
\end{figure}

\subsection{RQ1: The Asymmetry Exists and Is Robust}

Table~\ref{tab:crossmodel} presents the aggregate results.
Despite receiving identical contradictory supervision during Phase~2, the text pathway (T2T) suffers significantly greater degradation than the audio pathway (A2T) across every model and seed.
This holds across fundamentally different architectural paradigms---early-fusion projector (SALMONN), deep-fusion integration (AF3), and native multimodal backbone (Qwen2.5-Omni)---with the generation-space gap $\Delta D^{\text{gen}}$ ranging from $+0.029$ to $+0.106$ and all bootstrap confidence intervals excluding zero.
At the latent level the asymmetry is even more pronounced: logit-space RRS reaches up to $+0.423$, indicating that the text pathway loses over 40\% more predictive probability mass for the original target than the audio pathway.
Three Qwen2-Audio seeds yield near-identical gaps ($\Delta D^{\text{gen}} \approx +0.054$), demonstrating stability across random initialization.

\noindent$\bullet$ \textbf{Statistical significance.}
Across all configurations, one-tailed Wilcoxon signed-rank tests yield $p$-values $< 0.001$ with Cohen's $d$ ranging from 0.213 to 0.517.
Bootstrap 95\% confidence intervals for $\Delta D^{\text{gen}}$ and both RRS remain strictly above zero in every run.

\begin{table*}[t]
\centering
\caption{Controlled analyses and structural probes (RQ2). Five candidate explanations are tested; all are ruled out as sufficient
explanations. All rows report $\Delta D = D_{\text{T2T}} - D_{\text{A2T}}$ unless noted otherwise. Significance: {$^{***}$}\,$p<0.001$;
{$^{**}$}\,$p<0.01$; {$^{\text{n.s.}}$}\,$p \geq 0.05$.}
\vspace{-3mm}
\label{tab:rq2}
\small
\begin{tabular*}{\textwidth}{@{\extracolsep{\fill}}l ccc ccc}
\toprule
& \multicolumn{3}{c}{\textbf{Generation Space}} & \multicolumn{3}{c}{\textbf{Logit Space}} \\
\cmidrule(lr){2-4} \cmidrule(lr){5-7}
Condition & $\Delta D$$\uparrow$ & RRS$\uparrow$ & $d$$\uparrow$ & $\Delta D$$\uparrow$ & RRS$\uparrow$ & $d$$\uparrow$ \\
\midrule
\textbf{Baseline}$^{\ddagger}$ & \textbf{$+$0.054} & $+$0.187 & 0.289$^{***}$  & \textbf{$+$0.315} & $+$0.402 & 0.546$^{***}$ \\
\midrule
\rowcolor{gray!12}
\multicolumn{7}{l}{\textit{(a) Direct text overwrite: pressure-modality variation}} \\
\quad Audio-only               & \textbf{$+$0.032} & $+$0.122 & 0.162$^{***}$  & \textbf{$+$0.152} & $+$0.558 & 0.416$^{***}$ \\
\quad Text-only                & \textbf{$+$0.071} & $+$0.240 & 0.367$^{***}$  & \textbf{$+$0.450} & $+$0.863 & 0.854$^{***}$ \\
\midrule
\rowcolor{gray!12}
\multicolumn{7}{l}{\textit{(b) Projector shielding: projector perturbation with decoder noise control}} \\
\quad Projector $\sigma=0.50$  & \textbf{$+$0.055} & $+$0.192 & 0.297$^{***}$  & \textbf{$+$0.318} & $+$0.409 & 0.548$^{***}$ \\
\quad Projector $\sigma=0.75$  & \textbf{$+$0.059} & $+$0.211 & 0.321$^{***}$  & \textbf{$+$0.321} & $+$0.416 & 0.564$^{***}$ \\
\quad Projector $\sigma=1.00$  & \textbf{$+$0.055} & $+$0.190 & 0.297$^{***}$  & \textbf{$+$0.313} & $+$0.396 & 0.543$^{***}$ \\
\quad Decoder noise ctrl $\sigma=0.50$ & \textbf{$+$0.039} & $+$0.138 & 0.223$^{***}$ & \textbf{$+$0.330} & $+$0.415 & 0.567$^{***}$ \\
\midrule
\rowcolor{gray!12}
\multicolumn{7}{l}{\textit{(c) Contradictory replacement (MELD); overall $D_T^{\text{gen}} \approx 0.50$ vs.\ baseline $0.29$}} \\
\quad A2T-fair training        & \textbf{$+$0.058} & $+$0.116 & 0.385$^{***}$  & \textbf{$-$0.002} & $-$0.000 & $-$0.004$^{\text{n.s.}}$ \\
\quad T2T-fair training        & \textbf{$+$0.050} & $+$0.099 & 0.323$^{***}$  & \textbf{$+$0.046} & $+$0.005 & 0.088$^{\text{n.s.}}$ \\
\midrule
\rowcolor{gray!12}
\multicolumn{7}{l}{\textit{(d) Routing depth protection$^{\dagger}$; $\Delta D = D_{\text{routed}} - D_{\text{direct}}$}} \\
\quad TTS-routed (full audio stack)     & \textbf{$-$0.010} & $-$0.034 & $-$0.065$^{\text{n.s.}}$ & \textbf{$-$0.081} & $-$0.097 & $-$0.264$^{***}$ \\
\quad Text-routed (extra linear block)  & \textbf{$-$0.002} & $-$0.007 & $-$0.027$^{\text{n.s.}}$ & \textbf{$+$0.012} & $+$0.037 & $+$0.151$^{**}$ \\
\midrule
\rowcolor{gray!12}
\multicolumn{7}{l}{\textit{(e) Layer-specific gradient pressure: decoder layer freezing during Phase~2}} \\
\quad Freeze layers 0--8      & \textbf{$+$0.064} & $+$0.219 & 0.342$^{***}$  & \textbf{$+$0.356} & $+$0.653 & 0.627$^{***}$ \\
\quad Freeze layers 10--29    & \textbf{$+$0.057} & $+$0.190 & 0.293$^{***}$  & \textbf{$+$0.350} & $+$0.529 & 0.610$^{***}$ \\
\bottomrule
\multicolumn{7}{l}{\footnotesize $^{\ddagger}$\,Qwen2-Audio both-pathway contradictory pressure.
$^{\dagger}$\,Phase~1 acquisition matched: TTS 0.777/0.777; text-routed 0.772/0.773.} \\
\end{tabular*}
\vspace{-3mm}
\end{table*}

\noindent$\bullet$ \textbf{Temporal trajectory.}
Figure~\ref{fig:metric_heatmap} shows that pathway divergence emerges within the first few hundred steps and remains directionally positive throughout training.
The gap magnitude is not strictly monotonic: it can fluctuate and partially narrow at later steps in some models (notably Qwen2.5-Omni after step~1000), while preserving the same overall direction.

\noindent$\bullet$ \textbf{Acquisition confound.}
The ANCOVA pathway coefficient $\beta_{\text{path}}$ is positive and significant for reported seeds ($\beta_{\text{path}} = +0.057$, $p = 1.67 \times 10^{-4}$ for seed$_{42}$; $p = 1.41 \times 10^{-4}$ for seed$_{43}$; see Eq.~\ref{eq:ancova}), confirming that the asymmetry persists after controlling for differential Phase~1 acquisition strength.

\noindent$\bullet$ \textbf{Category and metric robustness.}
The asymmetry is directionally positive across all five retained question categories ($n \geq 30$).
Excluding the strongest category (sound-content, $d = 0.81$), the residual aggregate effect remains significant across all three seeds ($d$: 0.146--0.193, $p < 0.003$), ruling out a single-category explanation.
Figure~\ref{fig:metric_heatmap} confirms that across all four models and both semantic spaces, the text pathway shows consistently larger relative declines.
A leakage sanity check finds no significant correlation between per-item semantic overlap and per-item asymmetry (Spearman $\rho = 0.064$, $p = 0.528$).
Qualitatively, the audio pathway preserves the perceptual structure of the original content after Phase~2, while the text pathway drifts toward plausible but content-mismatched substitutes.
\par\vskip-0.5cm\relax

\subsection{RQ2: What Drives the Asymmetry?}
\label{sec:rq2}

Having established that the asymmetry is real and robust, we systematically test five candidate explanations and find that none is sufficient on its own.
Table~\ref{tab:rq2} reports all controlled conditions against a common baseline.

\noindent$\bullet$ \textbf{(a) Direct text overwrite.}
\label{sec:controlled}
If text-pathway knowledge forgets more only because Phase~2 directly rewrites text-side representations, then removing text-side pressure should eliminate the gap.
Under audio-only pressure, the text pathway is never directly overwritten, yet T2T still forgets more than A2T in both generation and logit space (Table~\ref{tab:rq2}, block~(a)), ruling out a direct-overwrite-only explanation.

\noindent$\bullet$ \textbf{(b) Projector shielding.}
If the audio pathway forgets less only because its projected representation occupies a geometrically protected region, then perturbing the projector should reduce the gap.
We inject Gaussian noise into projector weights before Phase~2: $P' = P + \sigma Z$ ($Z \sim \mathcal{N}(0, I)$, $\sigma \in \{0.50, 0.75, 1.00\}$).
Noise-only perturbation causes negligible Phase~1 retention damage ($< 0.01$), confirming that the perturbation alters projector geometry without collapsing acquisition.
Yet the asymmetry remains positive throughout (Table~\ref{tab:rq2}, block~(b)), indicating that projector geometry alone does not explain the effect.
A decoder-noise control ($\sigma = 0.50$) confirms this is not a generic insensitivity to perturbation: it reduces $\Delta D^{\text{gen}}$ from $+0.054$ to $+0.039$.

\noindent$\bullet$ \textbf{(c) Contradictory replacement.}
\label{sec:meld}
The preceding controls retain the core mechanism of contradictory overwriting.
To test whether the asymmetry depends on this mechanism, we replace Phase~2 with a MELD-based emotion classification task~\cite{poria2019meld}: a correct-label, cross-domain new task with no contradictory pressure.
Each utterance provides paired audio and transcript with a shared emotion label, preserving pathway-matched comparability; both arms are evaluated on the original MusicQA paired test set to measure retention of musical knowledge.
The generation-space asymmetry persists under both arms ($\Delta D^{\text{gen}} = +0.058$ and $+0.050$; Table~\ref{tab:rq2}, block~(c)), and the absolute gap remains in the same band as the contradictory baseline ($\Delta D^{\text{gen}} \approx +0.054$) despite overall forgetting increasing substantially ($D_T^{\text{gen}} \approx 0.50$ vs.\ $0.29$).
In logit space, neither arm shows a significant asymmetry (both $p > 0.6$), indicating that the cross-domain shift induces large but approximately pathway-symmetric logit degradation.
Under this interference regime, the generation-space readout remains the clearer indicator of pathway-dependent forgetting.
\begin{figure}[t]
\centering
\includegraphics[width=\linewidth]{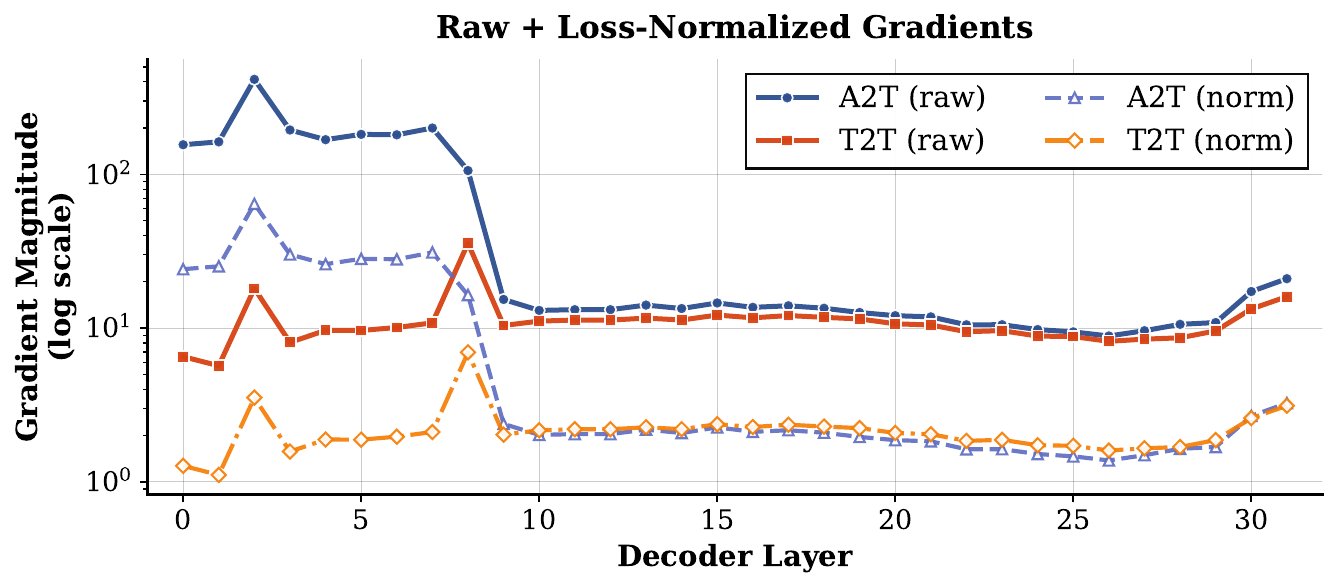}
\Description{Gradient analysis comparing A2T and T2T update magnitudes across Qwen2-Audio decoder layers during Phase 2.}
\par\vskip-5mm\relax
\caption{Exploratory gradient analysis on Qwen2-Audio (Phase~2).
}
\label{fig:gradients}
\par\vskip-0.5cm\relax
\end{figure}

\noindent$\bullet$ \textbf{(d) Routing depth protection.}
\label{sec:disentangle}
The audio pathway traverses additional encoder and projector stages, so the asymmetry could reflect routing depth rather than pathway identity.
We address this with two complementary probes (Table~\ref{tab:rq2}, block~(d)).
\emph{TTS-routed probe}: identical text descriptions are converted to synthetic speech and fed through the full audio stack, so that content is held constant while routing depth varies.
With Phase~1 acquisition matched, the routing-depth effect is non-significant in generation space ($d = -0.065$, $p = 0.253$) but significant in logit space ($d = -0.264$, $p = 7.7 \times 10^{-7}$); against the primary asymmetry (generation $d = 0.289$, logit $d = 0.546$), depth accounts for at most a minor fraction of the gap.
\emph{Text-routed probe}: an extra linear projection block (Linear $4096 \to 4096$ with residual connection) is applied to all token embeddings via a forward hook on the embedding layer, adding a routing stage while keeping the input modality fixed as text.\footnote{The hook operates on all tokens including those generated autoregressively, and the training mixture includes general instruction data alongside paired music items, matching the primary Phase~1 recipe.}
Phase~1 acquisition is again matched ($0.772$ vs.\ $0.773$), and the generation-space forgetting gap is negligible ($\Delta D^{\text{gen}} = -0.002$, $d = -0.027$, n.s.); in logit space the extra block slightly \emph{increases} forgetting ($\Delta D^{\text{logit}} = +0.012$, $d = +0.151$, $p < 0.01$), the opposite of a protective effect.
The two probes have orthogonal confound structures---one varies modality while preserving the real audio stack, the other preserves modality while adding an extra block---yet converge on the same conclusion: routing depth alone does not account for the observed asymmetry.

\noindent$\bullet$ \textbf{(e) Layer-specific gradient pressure.}
\label{sec:mechsignal}
Loss-normalized gradient profiles during Phase~2 (Figure~\ref{fig:gradients}) show that A2T updates concentrate in early decoder layers (0--8) while T2T updates are relatively larger in layers 10--29.
If this pattern were causally explanatory, freezing the T2T-dominant band should substantially reduce $\Delta D$, while freezing the A2T-dominant band should increase it.
Neither prediction holds: $\Delta D^{\text{gen}} = +0.064$ under freezing layers 0--8 and $+0.057$ under freezing layers 10--29, both comparable to the unfrozen baseline ($+0.054$; Table~\ref{tab:rq2}, block~(e)).
The logit-space effect sizes under both conditions ($d = 0.627$ and $0.610$) exceed the baseline ($d = 0.546$), ruling out a simple layer-specific causal account: the asymmetry is a distributed, architecture-level property.

The effect survives all five controlled tests, consistent with a structural property of how these models retain knowledge along different input pathways.

\begin{table}[!t]
\centering
\caption{Replay-based probe on Qwen2-Audio (generation space). Replay reduces overall forgetting but does not materially narrow the T2T--A2T gap.}
\par\vskip-3mm\relax
\label{tab:replay}
\resizebox{\columnwidth}{!}{%
\begin{tabular}{lcccccc}
\toprule
Condition & $D_{\text{A2T}}$$\downarrow$ & $D_{\text{T2T}}$$\downarrow$ & $\Delta D$$\uparrow$ & RRS$\uparrow$ & $d$$\uparrow$ & $p$$\downarrow$ \\
\midrule
Baseline & 0.235 & 0.289 & \textbf{+0.054} & +0.187 & 0.289 & 1.3e-09 \\
Balanced 5\% & 0.179 & 0.240 & \textbf{+0.062} & +0.256 & 0.290 & 2.0e-11 \\
T2T-targeted 5\% & 0.171 & 0.225 & \textbf{+0.054} & +0.239 & 0.262 & 6.7e-09 \\
Balanced 10\% & \textbf{0.160} & \textbf{0.211} & \textbf{+0.051} & +0.243 & \textbf{0.245} & 6.4e-09 \\
\bottomrule
\end{tabular}
}% end resizebo
\par\vskip-0.3cm\relax
\end{table}

%%
%% ─── RQ3 ───────────────────────────────────────────────────────────────────
%%

\subsection{RQ3: Can the Asymmetry Be Mitigated?}

As shown in Table~\ref{tab:replay}, three replay conditions reduce overall forgetting for both pathways, but the absolute pathway gap $\Delta D$ remains in the same band as the baseline ($+0.051$--$+0.062$ versus $+0.054$) rather than collapsing toward zero.
The persistence of $\Delta D$ under replay suggests that text-pathway vulnerability reflects a structural property of how these models organize knowledge along different pathways, rather than a shallow training artifact amenable to simple intervention, ruling out the last candidate explanation~(f).

%%
%% ─── QUALITATIVE ───────────────────────────────────────────────────────────
%%

\begin{figure}[t]
\centering
\includegraphics[width=0.95\linewidth]{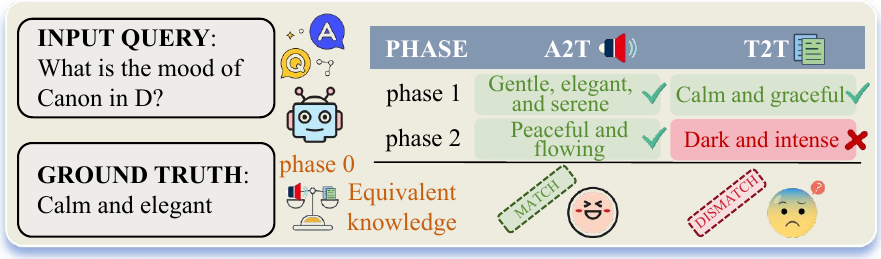}
\Description{Case study showing that A2T remains close to the correct mood after Phase 2 while T2T drifts to an incorrect mood.}
\par\vskip-3mm\relax
\caption{After Phase~2, A2T vs. T2T.}
\label{fig:casestudy}
\par\vskip-1mm\relax
\end{figure}
\subsection{Case Study}
Figure~\ref{fig:casestudy} illustrates a representative case from the music QA evaluation.
After Phase~1, both pathways correctly capture the calm, elegant character of \emph{Canon in D}.
After Phase~2, the audio pathway remains semantically anchored (``Peaceful and flowing''), while the text pathway drifts to a plausible but incorrect substitute (``Dark and intense''), suggesting that contradictory pressure overwrites content-specific bindings in the text pathway while leaving the audio pathway's perceptual grounding largely intact.

% \begin{table}[t]
% \centering
% \caption{Representative examples. After Phase~2, the audio pathway remains anchored to the original content, while the text pathway drifts toward plausible but content-mismatched substitutes.}
% \label{tab:casestudy}
% \small
% \begin{tabular}{p{2.1cm}p{1.5cm}p{1.7cm}p{1.7cm}}
% \toprule
% Question & Reference & A2T ($k{=}2$) & T2T ($k{=}2$) \\
% \midrule
% What sounds are present in this piece? & Guitar, drums, bass guitar & Guitar, bass, light percussion & Electronic beats, synth pads \\
% \midrule
% Describe the primary instruments you hear. & Piano and string ensemble & Piano, violin section & Brass section, drum kit \\
% \midrule
% What do you hear in the background? & Crowd applause and ambient noise & Audience clapping, room tone & Reverb, wind instruments \\
% \bottomrule
% \end{tabular}
% \end{table}

\section{Discussion}
\label{sec:discussion}

\subsection{Scope of the Present Findings}
\label{sec:entanglement}

In the music domain, the asymmetry is robust across architectures, models, metrics, interference regimes, and controlled probes (Section~\ref{sec:experiments}), providing evidence against the \emph{Pathway-Invariant Assumption}: retention in the studied models depends not only on what was learned, but on how that knowledge entered the model.
Our primary claim is scoped to the encoder--projector--LLM architectural family, whose categorical attributes (genre, instrumentation, mood, texture) enable strict pathway-level alignment; whether the observed direction generalizes to other modality pairs or architectural families remains open.

\noindent$\bullet$ \textbf{Entanglement.}
In all evaluated models, the audio pathway traverses additional encoder and projector stages, so pathway identity and routing depth are structurally entangled.
Two complementary probes (Section~\ref{sec:disentangle}) address this: the TTS-routed probe varies depth while changing modality, and the text-routed probe adds depth while holding modality fixed.
Neither finds a significant routing-depth effect at the generation level, and both are more consistent with pathway identity and input representation as dominant contributors.
A residual caveat is that the TTS probe's in-distribution Whisper processing and the text-routed probe's single linear block may underestimate depth effects relative to a full pretrained encoder; complete decomposition remains an open direction.

\noindent$\bullet$ \textbf{Perceptual grounding.}
An alternative account is that audio-pathway knowledge resists forgetting because perceptual grounding provides an implicit anchor.
The MELD control (Section~\ref{sec:meld}) bears directly on this account: without contradictory replacement, the generation-space gap $\Delta D^{\text{gen}}$ remains clearly positive, weakening the claim that perceptual grounding alone is sufficient.

\subsection{Open Problems and Broader Implications}

\noindent$\bullet$ \textbf{Pathway-aware forgetting mitigation.}
Uniform replay budgets do not close the gap (Section~\ref{sec:experiments}), suggesting that effective strategies may need to allocate protective resources in proportion to each pathway's vulnerability.

\noindent$\bullet$ \textbf{Broader implications.}
If pathway-dependent forgetting extends beyond the current setting, it carries implications for multimodal architecture design (routing as a retention-shaping factor), model editing and unlearning~\cite{meng2022rome,mitchell2022memit} (pathway-specific interventions for complete fact removal), and continual learning (pathway-aware rather than uniform mitigation).

\section{Conclusion}

We asked whether the route through which musical knowledge enters an audio-language model shapes its vulnerability to later forgetting.
PPCP, a three-phase protocol designed to address both the Pathway-Invariant Assumption and the Symmetric Continuation Bias, reveals a consistent asymmetry: text-pathway knowledge is forgotten more than matched audio-pathway knowledge across multiple architecturally distinct models.
The effect survives six targeted controls, and two independent routing-depth probes confirm that architectural depth alone does not account for it.
These results suggest that retention depends not only on what a model learned but on how that knowledge entered the model, motivating acquisition route as a new analytical dimension for forgetting research and multimodal system design.

\bibliographystyle{ACM-Reference-Format}
\bibliography{references}

\clearpage
\appendix

\section{Training Hyperparameters}
\label{app:hyperparams}

Table~\ref{tab:hyperparams} lists the full training configuration for each model.
All models use bfloat16 mixed precision, batch size 1 with gradient accumulation 16 (effective batch size 16), and a maximum sequence length of 2048 tokens.

\begin{table}[ht!]
\centering
\caption{Training hyperparameters and infrastructure.}
\label{tab:hyperparams}
\resizebox{\columnwidth}{!}{%
\small
\begin{tabular}{llcccc}
\toprule
Model & Trainable component & P1 LR & P2 LR & Precision & GPU \\
\midrule
Qwen2-Audio~\cite{chu2024qwen2audio} & Full model & 2e-5 & 5e-5 & bf16 & H100 \\
SALMONN~\cite{tang2024salmonn} & Vicuna (enc.+QF frozen) & 2e-5 & 5e-5 & bf16 & H100 \\
Audio Flamingo~3~\cite{goel2025af3} & Full model & 2e-5 & 7e-5 & bf16 & H100 \\
Qwen2.5-Omni~\cite{xu2025qwen25omni} & Thinker module & 2e-5 & 5e-5 & bf16 & H100 \\
\bottomrule
\multicolumn{6}{l}{\footnotesize Batch 1, grad.\ acc.\ 16, max len.\ 2048. P1: 3 ep.; P2: 2 ep.; AdamW.} \\
\multicolumn{6}{l}{\footnotesize Seeds: 42/43/44 (Qwen2-Audio), 42 (others). 4$\times$ NVIDIA H100 80GB.}
\end{tabular}%
}
\end{table}

\section{Per-Category Statistics}
\label{app:category}

Table~\ref{tab:category-full} reports per-category forgetting asymmetry for all eight question categories on Qwen2-Audio (seed~42, generation space).
Categories with $n < 30$ are separated below the line.
All five categories with $n \geq 30$ show positive $\Delta D$ direction; of the three excluded categories, one (use\_case\_purpose) also shows positive direction.

\begin{table}[ht!]
\centering
\caption{Per-category forgetting asymmetry (Qwen2-Audio, seed~42, generation space). Categories above the line have $n \geq 30$; those below are excluded from the main analysis due to small sample size.}
\label{tab:category-full}
\small
\begin{tabular}{lrccc}
\toprule
Category & $n$ & $\Delta D$ & Cohen's $d$ & $p$ \\
\midrule
sound\_content & 101 & $+$0.162 & 0.782 & 5.1e-11 \\
genre\_style & 104 & $+$0.048 & 0.277 & 2.9e-4 \\
instrument\_presence & 31 & $+$0.062 & 0.362 & 0.092 \\
mood\_emotion & 79 & $+$0.010 & 0.070 & 0.305 \\
unclassified & 133 & $+$0.023 & 0.137 & 0.339 \\
\midrule
use\_case\_purpose & 23 & $+$0.044 & 0.269 & 0.099 \\
vocal\_performer & 20 & $-$0.049 & $-$0.209 & 0.733 \\
tempo\_rhythm & 9 & $-$0.025 & --- & --- \\
\bottomrule
\end{tabular}
\end{table}

\section{Metric Robustness: Full Data}
\label{app:metrics}

Table~\ref{tab:metric-full} reports the complete forgetting asymmetry across four evaluation metrics and all six runs.
All three semantic metrics (sentence-transformer, NLI, BERTScore) yield significant positive asymmetry in every run.
ROUGE-L reaches significance for SALMONN, AF3, and Qwen2.5-Omni but not for Qwen2-Audio.

\begin{table}[t]
\centering
\caption{Full metric robustness data. $D_A$/$D_T$: mean forgetting for A2T/T2T; $\Delta D = D_T - D_A$. Significance: $^{***}p<0.001$; $^{**}p<0.01$; n.s.\ $p \geq 0.05$.}
\label{tab:metric-full}
\resizebox{\columnwidth}{!}{%
\small
\begin{tabular}{ll ccccc}
\toprule
Run & Metric & $D_A$ & $D_T$ & $\Delta D$ & $d$ & $p$ \\
\midrule
\multirow{4}{*}{\shortstack[l]{Qwen2-Audio\\(s42)~\cite{chu2024qwen2audio}}}
& Sent.-trans. & .235 & .289 & $+$.054 & .289$^{***}$ & 1.3e-9 \\
& NLI & .092 & .283 & $+$.192 & .370$^{***}$ & 2.6e-13 \\
& BERTScore & .029 & .033 & $+$.004 & .121$^{**}$ & 2.6e-3 \\
& ROUGE-L & .193 & .192 & $-$.001 & $-$.005$^{\text{n.s.}}$ & .18 \\
\midrule
\multirow{4}{*}{\shortstack[l]{Qwen2-Audio\\(s43)}}
& Sent.-trans. & .233 & .287 & $+$.054 & .297$^{***}$ & 5.6e-10 \\
& NLI & .089 & .295 & $+$.207 & .401$^{***}$ & 2.6e-14 \\
& BERTScore & .029 & .032 & $+$.003 & .107$^{**}$ & 7.1e-3 \\
& ROUGE-L & .189 & .196 & $+$.007 & .033$^{\text{n.s.}}$ & .073 \\
\midrule
\multirow{4}{*}{\shortstack[l]{Qwen2-Audio\\(s44)}}
& Sent.-trans. & .228 & .281 & $+$.053 & .282$^{***}$ & 1.7e-9 \\
& NLI & .069 & .292 & $+$.223 & .431$^{***}$ & 4.0e-17 \\
& BERTScore & .029 & .033 & $+$.004 & .122$^{**}$ & 4.8e-3 \\
& ROUGE-L & .190 & .195 & $+$.005 & .026$^{\text{n.s.}}$ & .11 \\
\midrule
\multirow{4}{*}{SALMONN~\cite{tang2024salmonn}}
& Sent.-trans. & .232 & .276 & $+$.044 & .288$^{***}$ & 3.9e-9 \\
& NLI & .163 & .275 & $+$.113 & .237$^{***}$ & 3.3e-10 \\
& BERTScore & .024 & .029 & $+$.005 & .203$^{***}$ & 6.6e-7 \\
& ROUGE-L & .174 & .199 & $+$.026 & .150$^{***}$ & 4.3e-5 \\
\midrule
\multirow{4}{*}{\shortstack[l]{Audio Flamingo~3\\~\cite{goel2025af3}}}
& Sent.-trans. & .160 & .268 & $+$.109 & .524$^{***}$ & 3.2e-27 \\
& NLI & .090 & .287 & $+$.197 & .391$^{***}$ & 2.8e-16 \\
& BERTScore & .018 & .031 & $+$.012 & .391$^{***}$ & 1.5e-17 \\
& ROUGE-L & .153 & .203 & $+$.050 & .250$^{***}$ & 2.9e-11 \\
\midrule
\multirow{4}{*}{\shortstack[l]{Qwen2.5-Omni\\~\cite{xu2025qwen25omni}}}
& Sent.-trans. & .242 & .271 & $+$.029 & .213$^{***}$ & 2.1e-5 \\
& NLI & .162 & .257 & $+$.096 & .225$^{***}$ & 8.6e-7 \\
& BERTScore & .028 & .031 & $+$.003 & .120$^{**}$ & 5.4e-3 \\
& ROUGE-L & .171 & .194 & $+$.023 & .142$^{**}$ & 1.9e-3 \\
\bottomrule
\end{tabular}%
}
\end{table}

\section{Additional Control Details}
\label{app:controls}

\subsection{Sentence-Form MELD Control}
\label{app:meld_sentence}

The MELD control replaces contradictory Phase~2 supervision with a correct-label cross-domain task built from paired speech utterances and transcripts~\cite{poria2019meld}.
Each arm uses 12{,}000 Phase~2 items and is evaluated on the original 500-item MusicQA paired test set.
To keep the supervision closer to a generation task, emotion labels are expressed as short natural-language sentences rather than bare class tokens.

\begin{table}[ht!]
\centering
\caption{Sentence-form MELD control on Qwen2-Audio (seed~42). The generation-space asymmetry remains positive under both fair-training arms, while logit-space asymmetry stays weak and individually non-significant.}
\label{tab:meld-sentence-appendix}
\small
\begin{tabular}{lccccc}
\toprule
Condition & $D_A^{\text{gen}}$ & $D_T^{\text{gen}}$ & $\Delta D^{\text{gen}}$ & $d^{\text{gen}}$ & $p^{\text{gen}}$ \\
\midrule
A2T-fair & 0.443 & 0.501 & $+$0.058 & 0.385 & $< 0.001$ \\
T2T-fair & 0.455 & 0.505 & $+$0.050 & 0.323 & $< 0.001$ \\
\midrule
Condition & \multicolumn{2}{c}{$\Delta D^{\text{logit}}$} & \multicolumn{2}{c}{$d^{\text{logit}}$} & $p^{\text{logit}}$ \\
\midrule
A2T-fair & \multicolumn{2}{c}{$-$0.002} & \multicolumn{2}{c}{$-$0.004} & n.s. \\
T2T-fair & \multicolumn{2}{c}{$+$0.046} & \multicolumn{2}{c}{0.088} & n.s. \\
\bottomrule
\end{tabular}
\end{table}

Relative to the contradictory baseline ($\Delta D^{\text{gen}} \approx +0.054$), sentence-form MELD increases overall forgetting while keeping the absolute pathway gap in the same band.
This strengthens the conclusion that the main asymmetry is not specific to contradictory replacement.

\subsection{TTS-Routed Probe}
\label{app:ttsprobe}

The TTS-routed probe feeds the same canonical text descriptions through two routes: a direct text route and a routed audio route obtained by first synthesizing the description into speech and then processing it through the full audio stack.
This keeps the semantic content fixed while changing the route structure.
Phase~1 acquisition is tightly matched in generation space ($0.7774$ direct vs.\ $0.7768$ routed).

\begin{table}[ht!]
\centering
\caption{TTS-routed probe on Qwen2-Audio (seed~42). Routed audio yields only a modest retention advantage relative to the direct text route.}
\label{tab:ttsprobe-appendix}
\resizebox{\columnwidth}{!}{%
\small
\begin{tabular}{lcccccc}
\toprule
Route & P1 gen & P2 gen & $D^{\text{gen}}$ & P1 logit & P2 logit & $D^{\text{logit}}$ \\
\midrule
Direct text & 0.7774 & 0.4751 & 0.3023 & $-$1.1741 & $-$2.0079 & 0.8338 \\
TTS-routed & 0.7768 & 0.4847 & 0.2921 & $-$1.2887 & $-$2.0412 & 0.7526 \\
\bottomrule
\end{tabular}
}
\end{table}

The routed path forgets slightly less in generation space ($\Delta D^{\text{gen}} = -0.010$, $d = -0.065$, n.s.) and more clearly less in logit space ($\Delta D^{\text{logit}} = -0.081$, $d = -0.264$, $p < 0.001$), indicating at most a modest route-structure contribution relative to the primary asymmetry.

\subsection{Text-Routed Probe: Implementation and Validation}
\label{app:textrouted}
The text-routed probe adds a single linear projection block (Linear $4096 \to 4096$) with a residual connection to the text pathway via a forward hook on the embedding layer.
The block is initialized to exact zeros so that the residual connection acts as an identity at initialization; this ensures that Phase~0 scores for the direct and routed pathways are identical.

\noindent$\bullet$ \textbf{Phase~0 validation.}
Table~\ref{tab:textrouted-phases} confirms that the zero-initialized probe produces exact score parity at Phase~0 (gap $= 0.000000$ in both spaces), validating the implementation.
Phase~1 acquisition is matched to within $0.0004$ in generation space, satisfying acquisition comparability.

\noindent$\bullet$ \textbf{Design choices.}
Two alternative implementations were tested and rejected during development.
Xavier-normal initialization (gain $= 0.1$) produced negligible logit-space deviation but caused generation-space scores to diverge by $0.31$ due to autoregressive error accumulation.
Passing routed embeddings via \texttt{inputs\_embeds} instead of \texttt{input\_ids} caused KV-cache inconsistencies in HuggingFace's \texttt{generate()}, producing a $0.31$ generation gap even with zero-valued embeddings.
Both issues were resolved by the zero-initialized forward-hook design.

\begin{table}[ht!]
\centering
\caption{Text-routed probe: scores at each PPCP phase.}
\label{tab:textrouted-phases}
\small
\begin{tabular}{lcccc}
\toprule
Phase & Direct (gen) & Routed (gen) & Direct (logit) & Routed (logit) \\
\midrule
0 & 0.7289 & 0.7289 & $-$2.9010 & $-$2.9010 \\
1 & 0.7724 & 0.7728 & $-$3.0163 & $-$3.0182 \\
2 & 0.4543 & 0.4568 & $-$3.3507 & $-$3.3650 \\
\bottomrule
\end{tabular}
\end{table}

These phase scores imply generation-space forgetting of $0.318$ for the direct path and $0.316$ for the routed path ($\Delta D^{\text{gen}} = -0.002$, $d = -0.027$, n.s.), while in logit space the extra block slightly increases forgetting ($0.335 \rightarrow 0.347$, $\Delta D^{\text{logit}} = +0.012$, $d = +0.151$, $p < 0.01$).
Taken together, the text-routed probe does not reproduce the main protective effect.

\section{Phase~0 and Phase~1 Baseline Scores}
\label{app:baselines}

Table~\ref{tab:baselines} reports the raw pathway scores before any forgetting pressure is applied.
Phase~0 reflects pretrained model performance; Phase~1 reflects post-injection performance after matched pathway activation.
Acquisition gain ($G = S^{(1)} - S^{(0)}$) is positive for both pathways in all models, confirming that Phase~1 successfully engages both pathways on the target knowledge.

\begin{table}[ht!]
\centering
\caption{Phase~0 and Phase~1 scores and acquisition gains (seed~42). Gen: generation-space similarity; Log: mean log-prob.}
\label{tab:baselines}
\resizebox{\columnwidth}{!}{%
\small
\begin{tabular}{l cccc cc}
\toprule
& \multicolumn{2}{c}{\textbf{Phase~0 (gen)}} & \multicolumn{2}{c}{\textbf{Phase~1 (gen)}} & \multicolumn{2}{c}{\textbf{Gain}} \\
\cmidrule(lr){2-3} \cmidrule(lr){4-5} \cmidrule(lr){6-7}
Model & $S_A$ & $S_T$ & $S_A$ & $S_T$ & $G_A$ & $G_T$ \\
\midrule
Qwen2-Audio~\cite{chu2024qwen2audio} & 0.643 & 0.715 & 0.719 & 0.773 & $+$0.075 & $+$0.057 \\
SALMONN~\cite{tang2024salmonn} & 0.629 & 0.739 & 0.746 & 0.790 & $+$0.117 & $+$0.051 \\
Audio Flamingo~3~\cite{goel2025af3} & 0.646 & 0.755 & 0.692 & 0.795 & $+$0.046 & $+$0.041 \\
Qwen2.5-Omni~\cite{xu2025qwen25omni} & 0.576 & 0.645 & 0.749 & 0.777 & $+$0.173 & $+$0.132 \\
\bottomrule
\end{tabular}%
}
\end{table}

\end{document}